\documentclass[prc,preprint]{revtex4}

\def\pt{p_T}
\def\ppt{$p_T$}
\def\ep{$(\eta,\phi)$}
\def\bq{ \begin{eqnarray}}
\def\eq{\end{eqnarray}}
\def\dpt{$\Delta p_T$}
\def\fq{$F_q$}
\def\rh{$\rho_0$}
\def\ee{Eq.\ }
\begin{document}


\title{Recognizing Critical Behavior amidst Minijets at the \\ Large Hadron Collider}
\author{Rudolph C. Hwa}
\affiliation
{Institute of Theoretical Science and Department of
Physics\\ University of Oregon, Eugene, OR 97403-5203, USA}

\date{November 2014}

\begin{abstract}

The transition from quarks to hadrons in a heavy-ion collision at high energy is usually studied in two different contexts that involve very different transverse scales: local and non-local. Models that are concerned with the \ppt\ spectra and azimuthal anisotropy belong to the former, i.e., hadronization at a local point in \ep\ space, such as  the recombination model. The non-local problem has to do with quark-hadron phase transition  where collective behavior through near-neighbor interaction can generate patterns of varying sizes in the \ep\ space. The two types of problems are put together in this paper both as brief reviews separately and to discuss how they are related to each other. In particular, we ask how minijets produced at LHC can affect the investigation of multiplicity fluctuations as signals of critical behavior. It is suggested that the existing data from LHC have sufficient multiplicities in small \ppt\ intervals to make feasible the observation of distinctive features of clustering of soft particles, as well as voids, that characterize the critical behavior at phase transition from quarks to hadrons, without any ambiguity posed by the clustering of jet particles.
\end{abstract}
\maketitle

\section{Introduction}
Critical phenomenon is a subject that has been extensively investigated in many fields, but never explicitly identified in heavy-ion collisions. There are two types of phase transitions involved when nuclei collide. One is QCD deconfinement caused by compression of the nuclei such that quarks are liberated from the nucleons, as probed in the Beam Energy Scan (BES) program at RHIC for collision energy ranging from 7.7 to 19.6 GeV \cite{lk}. The other is quark-hadron phase transition for a quark-gluon plasma that has previously been hot and dense after creation by nuclear collisions at very high energy, but has expanded enough so that confinement forces set in at the end of evolution to form hadrons. One expects such a phase transition (PT) to be operative at energy ~200GeV at RHIC, but no definitive signal has been reported. A baryon-free central region suitable for the Ginzburg-Landau description of PT \cite{h1} is more likely to be created and to offer detectable signature only at the Large Hadron Collider (LHC) in Pb-Pb collisions. Our concern in this paper is exclusively on the latter type of PT, where hadronization is a process far removed and distinct from the initial configurations of colliding nuclei. However, at LHC there is the added complication of jets and shower partons that may contaminate the signals for PT. It is our objective here to clarify these various related issues.

	The observables in a heavy-ion collision are the momentum variables of hadrons in $(\eta,\phi,p_T)$,  apart from their identities. Since the partons in the medium are not directly observable, one must infer from the multiparticle distribution in $(\eta,\phi,p_T)$ what the nature of the dynamical system is before hadronization occurs. Just as the medium exists over a period of time, so also does one expect the formation of hadrons to occur at various times. The experiment collects all the particles produced in its acceptance windows irrespective of the time of emission in each event. After averaging over many events, it is hard to recognize from the inclusive distributions or even particle correlations whether the quark-hadron transition is of a type characteristic of critical behavior.

	Hydrodynamical models take the macroscopic approach in describing the evolution of the medium \cite{ph}. In treating the fluid flow of the energy-momentum tensor that averages over the microscopic variables of the constituents, and adopting simple schemes for freeze-out, the approach relinquishes any interest in the question of the collective behavior of the underlying quarks. That is not to say that there is no collective motion of the fluid. The central theme of our study here is the critical behavior that arises from the tension between the ordered (collective) and the disordered (thermal) dynamics. The macroscopic variables used in hydro models can describe collective flow, but not near-neighbor interactions that generate cooperative behavior.

	Event generators that incorporate microscopic dynamics such as parton scattering and hadronization by coalescence, as in A Multi-Phase Transport (AMPT) model \cite{bz}, do not treat the movements of large patches of partons in reaction to confinement forces, so they also do not contain the dynamics of phase transition. They do reproduce the data on hadronic spectra in  much wider ranges in phase space than hydro models.

	Since the systems created in heavy-ion collisions are so complex, any theoretical treatment of them must rely on some approximations and some degree of averaging that can render a tractable description of what is regarded as essential. The focus on \ppt\ distributions and azimuthal anisotropy is orthogonal to the issues of critical behavior, although both sides of the landscape involve the interface of quarks and hadrons. 

	To extract the signature of quark-hadron PT from the massive data acquired in heavy-ion collisions, it is essential that averaging is not done that erases the signal from the start. Even before averaging, the nature of the structure in momentum space where the thousands of produced particles populate in one event may contain overlapping features that have origins from different dynamics. Clustering due to collective behavior may at some level appear to be similar to clusters of fragmentation products of hard jets. To be able to distinguish the different features is crucial to the construction of a program that can be successful in discovering any evidence for or against critical behavior in heavy-ion collisions.

	A good measure of criticality for heavy-ion collisions was proposed two decades ago \cite{hn} on the basis of Ginzburg-Landau theory of second-order PT \cite{gl}. It emphasizes the study of the scaling properties of multiplicity fluctuations over a wide range of bin sizes. The numerical value of a scaling exponent $\nu$ was derived analytically. Although the universality of $\nu$ was verified in a laser experiment at the onset of lasing \cite{yo}, it has never been tested in heavy-ion experiments because of complications arising from insufficiently high multiplicities. Now, with the data available from LHC on Pb-Pb collisions at 2.76 TeV , dedicated analysis can begin. On the other hand, it has also been found in the study of \ppt\ spectra of hadrons at LHC that shower partons dominate over thermal partons because of profusely produced minijets at that high collision energy \cite{zh}. It is therefore important to understand how the shower partons affect the multiplicity fluctuations due to critical clustering. In clarifying those issues we shall review all the main features of each, and show why minijets present no problem in the search for scaling behavior, whether or not critical quark-hadron PT exists.

\section{Brief review of signature of critical behavior}

Unlike most problems in statistical physics where critical phenomena occur, temperature cannot be controlled in heavy-ion collisions. Thus, even the notion of criticality has no experimental relevance unless some phenomenological measure can be devised that can be shown at least on theoretical grounds to reveal the occurrence of quark-hadron PT. In a high-energy collision quarks turn into hadrons anyway whether or not there is a PT, as in $pp$ or leptonic collisions. Those are local properties where temperature is not an appropriate description of the system. To be qualified as a PT the system must have enough constituents to form a quasi-equilibrium state in which near-neighbor interactions can result in correlated behavior in patches of neighborhoods as in ferromagnetism or in QCD-confined clusters. Those are semi-global properties that require a spatially extended region to display their characteristics.

	In heavy-ion collisions the \ep\ components of a particle momentum can be related to a point on the surface of a cylinder that contains the quark-gluon plasma (QGP). Since the plasma lasts for some time, it is reasonable to expect, even without hydrodynamics, that the interior is hotter than the surface, so hadrons are more likely to be formed on the surface as the system expands and cools. It is the spatial pattern of where the hadrons are formed that is of interest. But those patterns at different times are superimposed on one another as time progresses, so at the end, when the detector collects all the particles in any given event, the characteristic features of the fluctuation patterns disappear in the cumulative result. Since it is not feasible to cut the time sequence into small $\Delta t$ intervals, the best that we can do is to make \dpt\ cuts in the hope that the high and intermediate \ppt\ particles are mostly emitted at early times and that  only the bulk matter that remains at late time undergoes more graduate PT at low \ppt. Lacking the ability to tune the temperature $T$, we need a measure that captures the essence of quark-hadron PT without reference to $T$.

	The physical basis for our interest in studying the fluctuations of spatial patterns is that it is the characteristic property of critical behavior to exhibit patches of all sizes \cite{bi}. At $T$ well below the critical temperature $T_c$, near-neighbor interactions dominate over thermal randomization so there is an ordered state throughout, as exemplified by the alignment of spins in a ferromagnet. At $T>T_c$ thermal agitation dominates and the state becomes disordered. At $T=T_c$ the tension between the two forces results in patches of ordered state of various sizes. For quark-hadron PT in heavy-ion collisions to exhibit that kind of behavior the lego plot of hadrons formed in the \ep\ space should show clusters of all sizes. Since there is no characteristic scale in the problem, there should be scaling property of the hadron multiplicities in bin sizes. Concrete models have been devised to simulate such critical behavior for hadron observables, ranging from the Ising model \cite{cgh} on the one end to contracting clusters due to color confinement \cite{hy12} on the other. Power-law behaviors have indeed been found with characteristic scaling exponents.

	To quantify the fluctuation properties of bin multiplicities, normalized factorial moments $F_q$ have been used, where $F_q$ is defined by \cite{bp}
\bq
F_q={f_q\over f_1^q} \ ,  \qquad \qquad  f_q=\sum_{n=q}^\infty {n!\over (n-q)!}\ P_n .   \label {1}
\eq
$P_n$ is the multiplicity distribution in a 2D bin of size $\delta^2$, so $f_q=\left< n(n-1)\cdots (n-q+1)\right>$. There are two ways of interpreting the average $\left< \cdots \right>$. One is to fix the location of the bin, average over all events, and study the result as a function of the bin width $\delta$. In that case $P_n(\delta)$ is the probability of having $n$ particles in that bin after $N_{\rm evt}$ events. Such an average is called vertical average, and can be denoted by $\left< \cdots \right>_v$. The other way is to partition the 2D space into $M$ square bins (each of width $\delta$ on each side) and to take the average over all bins in one event. That is called horizontal average $\left< \cdots \right>_h$, for which $P_n(M)$ is the probability of having $n$ particles in any of the $M$ bins for any particular event. If the space is large and uniform (which can be achieved by use of the cumulative variables), one expects on the basis of ergodic principle that the two ways of averaging lead to the same result. Clearly, doing both averages yields better statistics. Since the spatial pattern displays explicitly the event structure, it is best to take the horizontal average for $f_q$ first, and then the vertical average of $F_q$, i.e.,
\bq
F_q=\left<{\left<n(n-1)\cdots(n-q+1\right>_h\over \left<n\right>_h^q}\right>_v   .   \label{2}
\eq
The advantage of studying \fq\ is that it filters out statistical fluctuations, for which the reader is referred to the original article \cite{bp}, and later reviews \cite{h1,dk}.

	If \fq\ has a power-law dependence on $M$
\bq
F_q \propto M^{\varphi_q} \ ,   \label {3}
\eq
the phenomenon is referred to as intermittency \cite{bp}. It implies the lack of any particular spatial scale in the system. Various experiments have found intermittency in collisions of leptons, hadrons and nuclei with positive intermittency index $\varphi_q$ \cite{es,dk}. Since critical phenomenon like quark-hadron PT has scaling properties, one can expect intermittency to be indicative of the existence of QGP in nuclear collisions \cite{bh}. However, what is indicative is not a necessary condition. Hadronization of QGP may take various routes, and there is no consensus on what that route must necessarily be.

	To build a more solid foundation on quark-hadron PT, it is appealing to refer to the phenomenological theory of Ginzburg-Landau (GL) that has found universal validity in describing critical behaviors \cite{gl,bi}. The GL theory is a mean field theory and as such cannot predict reliably the critical exponents for systems with large fluctuations. However, it is sufficient for our purpose since the temperature $T$ is not subject to control in heavy-ion experiments.
The first question is how to relate the order parameter $\Phi$ in that theory to the observables in heavy-ion collisions. In \cite{hn}, $|\Phi|^2$ is identified with hadron density, and specific scaling behavior of \fq\ was found analytically. Two years later the same treatment was applied to nonlinear optics where the number of photons created at the threshold of lasing can be related to second-order PT \cite{yo}. There, the GL order parameter $\Phi$ is the complex field amplitude $E$ of the single-mode laser. Experimentally, it is possible to control the operating point of the laser in the same way that $T$ is under control in a condensed-matter system. It turns out that at the threshold of lasing exactly the same scaling behavior is found as predicted for quark-hadron PT. Regarding that as an experimental confirmation of the relevance of GL theory to the production of particles, photons or hadrons, let us now go back to the problem of heavy-ion collisions and describe what that theory predicts for multiplicity fluctuations if the QGP undergoes a quark-hadron PT.	

	The GL free-energy density is \cite{gl,bi} 
\bq
{\cal F}[\Phi] = a |\Phi({\bf r})|^2 + b |\Phi({\bf r})|^4 \ ,   \label {4}
\eq
 where a gradient term $\partial \Phi/\partial {\bf r}$ is usually also on the RHS, but will be neglected if we ignore the spatial dependence within a bin in the space \ep\ that we shall consider. We shall do only vertical average at one fixed bin, so according to our identification of $|\Phi|^2$ with hadron density, the average multiplicity in the bin of size $\delta^2$ is 
 \bq
\left< n \right> = \int_{\delta^2} d^2{ r}\ |\Phi({\bf r})|^2 = \delta^2 \Phi^2 .  \label{5}
\eq 
That means that hadrons are formed wherever $\Phi$ is nonzero for $a < 0$ and $b > 0$ in \ee(4), which corresponds to having a (hadron) condensate in a second-order PT. The case of a first-order PT with $a > 0, b < 0$ and a third $|\Phi|^6$ term in \ee(4) is more complicated, and will be commented on below after the completion of our discussion about the second-order PT.

	Since the system can fluctuate around the minimum of ${\cal F}[\Phi]$, the multiplicity distribution involves an integration over all $\Phi$
\bq
P_n = {1\over Z} \int {\cal D}\Phi\ P_n^0[\Phi]\ e^{-F[\Phi]} \ ,  \label{6}
\eq
where
\bq
F[\Phi] = \delta^2 {\cal F}[\Phi]\ ,  \qquad\qquad Z= \int {\cal D} \Phi\ e^{-F[\Phi]} \ ,  \label{7}
\eq
and $P_n^0[\Phi]$  is the Poisson distribution 
\bq
P_n^0[\Phi]={1\over n!} (\delta^2\Phi^2) e^{-\delta^2\Phi^2}  .   \label{8}
\eq
Applying \ee (6) to (1) one can obtain an analytical formula for \fq, whose dependence on $\delta$  is found not to satisfy a simple power-law description, as described by \ee(3) for intermittency \cite{hn}. However, it can be established that it has the property 
\bq
F_q \propto F_2^{\beta_q}   \label{9}
\eq
for an extended range of \fq, a behavior that has been referred to as $F$-scaling. Furthermore, the scaling exponent $\beta_q$ satisfies 
\bq
\beta_q = (q-1)^\nu \ ,  \qquad\qquad  \nu=1.304   \label{10}
\eq
to a high degree of accuracy. The index $\nu$ is independent of the GL parameters $a$ and $b$ (and therefore independent of $T$) so long as $a < 0$ and $b > 0$, and independent of the bin size $\delta$ \cite{hn,h2}. It is an amazingly simple numerical quantity that connects GL theory to hadronic observable in heavy-ion collisions. In photon counting at the threshold of lasing this index $\nu$ is verified accurately \cite{yo}. It is so far the only experiment that has examined the relevance of $F$-scaling to PT in any system.

	There are event generators that simulate particle production in heavy-ion collisions. They are usually tuned to reproduce the data on \ppt\ spectra and azimuthal anisotropy. Recently, the AMPT code \cite{bz} has been used to generate multiparticle production by Pb-Pb collisions at $\sqrt {s_{NN}}=2.76$ TeV and the multiplicity fluctuations have been examined using factorial moments \fq\ defined in \ee(2) \cite{gup}. It is found that not only is there no intermittency; but further, the exponents $\varphi_q$ in \ee(3) are negative for all $q\ge 2$. It means that there are no large fluctuations. The absence of any sign of PT is not surprising, since no collective interaction has been built into the code. The analysis done in \cite{gup} is, however, preparatory to the application of the method to the real data from LHC.

	Returning to theory, it should be noted that the GL theory of PT, being of mean field, is too smooth to be able to describe the fluctuation pattern in the whole \ep\ space of any event. The Ising model in 2D that simulates ferromagnetism on a square lattice is well suited to amend the deficiency of the GL theory on the one hand, and to make more direct connection with the event structure of particle production on the other. That problem has been investigated in Ref.\ \cite{cgh}. Hadron multiplicity in a cell of $\epsilon^2$ lattice sites is defined in proportion to the net spin of the cell along the direction of the net magnetization of the whole lattice, and is treated as zero multiplicity if the net spin is in the opposite direction. Thus each Ising configuration corresponds to an event with hadron occupancy fluctuating throughout the lattice. Since $T$ is under control in the Ising simulation, one can examine the $T$ dependence of the average multiplicity $\left< n\right>$ in all cells and determine the critical temperature $T_c$ so that $\left< n\right>$ is large for $T<T_c$ and small for $T>T_c$ with a sharp transition at $T_c$. This is a simulation of PT because the Ising Hamiltonian has near-neighbor interaction that favors parallel spin but is randomized by thermal motion in accordance to the Boltzmann factor. Thus the Ising model is orthogonal to event generators, like AMPT, which has parton scattering dynamics but not collective interaction.

	Since each event in a heavy-ion collision corresponds to an Ising configuration, the spatial pattern of the event structure can be analyzed by the horizontal factorial moments, and the vertical averaging can be achieved by simulating many Ising configurations at a fixed $T$. Thus it is possible to calculate \fq, defined in \ee(2), as a function of bin size $\delta^2$, with each bin consisting of $(\delta/\epsilon)^2$ cells. It is found that strict power-law behavior of intermittency defined by \ee(3) occurs only at $T=T_c$; however, $F$-scaling defined by \ee(9) is valid for a range of $T<T_c$, but not at $T>T_c$. The dependence of $\beta_q$ on $q$ given in \ee(10) is indeed verified by the Ising model, but  the index $\nu$ shows dependence on $T$ \cite{cgh}. It varies from $\nu=1.6$ at low $T$ to $\nu=1$ at $T=T_c$, so the average $\nu=1.3$ agrees well with the result from GL given in \ee(10). Thus we learn that if the Ising model is a valid description of quark-hadron PT and supplements the gross prediction of GL theory, then hadrons can form at a range of temperature after the quark system cools below the point when the constituents of hadrons can no longer remain deconfined.

	The above discussion is all about second-order PT. If the quark-hadron PT is first order, the GL free-energy density for that phenomenon corresponds to $a > 0, b < 0$ in \ee(4), so ${\cal F}[\Phi]$ must have a $c|\Phi({\bf r})|^6$ term, with $c > 0$, in order that $\exp(-F[\Phi])$ be bounded at large $|\Phi|$. Thus ${\cal F}[\Phi]$ has a maximum located at a value of $|\Phi|^2$ between 0 and the point where ${\cal F}[\Phi]$ has a minimum, which is where the system jumps to in a transition. The quantity that characterizes the strength of that transition is \cite{h3}
\bq
g = -\left( 1+{b\over 2 \sqrt{ac}} \right)  ,   \label{11}
\eq
which must be $> 0$ for ${\cal F}[\Phi]$ to be $< 0$ at the minimum. When $g$ is 0.6 or larger, the minimum is so deep compared to the bump at the maximum that the PT is very nearly second order. So the interesting region for first-order PT is for $0 < g < 0.6$, where small $g$ corresponds to strong first order. A study of the behavior of \fq\ has been carried out in Ref.\ \cite{h3}, where it is found that $F$-scaling (9) is well satisfied and that the index $\nu$ depends on $g$. It decreases from $\nu=1.45$ at $g = 0$ asymptotically to $\nu=1.30$ at $g \ge 0.6$. Since $\nu$ is a measurable quantity, it offers a simple procedure to characterize quark-hadron PT in the absence of any way to determine the GL parameters or the temperature.

From the results summarized above for first- and second-order PTs it is evident that the scaling exponent $\nu$ does not provide distinctive characterization of the nature of PT, except if its value is very nearly 1.3. If $\nu$ is larger than 1.3, it could mean that the quark-hadron system hadronizes at a $T$ lower than $T_c$ in second order, or supercooled in first order. In the latter case the system persists for some time in the deconfined phase despite the existence of a lower minimum in the GL free energy. On the other hand, it is also possible that other physical processes are at play to mask the expectations from the equilibrium dynamics of PT. Thus, we need other methods of analyses to exhibit a wider scope of the properties of hadronization.

	If a system exhibits intermittency as in \ee(3), or even if there is no strict power law but \fq\ increases with $M$, then the existence of large \fq\ implies large spatial fluctuations at small bin size $\delta$. To amplify the behavior at large \fq, it is natural to consider positive powers of \fq\ before vertical averaging. If we denote the quantity inside $\left< \cdots \right>_v$ in \ee(2) by $F_q^e$ for an event $e$, its fluctuation from its event average can be quantified by the double moment \cite{h4,ch}
\bq
C_{p,q}(M) = \left<\left[{F_q^e(M)\over \left< F_q(M)\right>_v}\right]^p\right>_v  \label {12}
\eq
where $p > 1$ need not be an integer. If $C_{p,q}(M)$ has a power-law behavior in $M$
\bq
C_{p,q}(M) \propto M^{\psi_q(p)}  ,   \label{13}
\eq
the phenomenon is referred to as erraticity. If further the exponent $\psi_q(p)$  depends linearly on $p$ in an interval of $p$ above 1, then the slope
\bq
\mu_q=d\psi_q(p)/dp   \label{14}
\eq
is referred to as erraticity index that is independent of $M$ and $p$. It was found that this way of measuring fluctuations can be applied to common problems in classical chaos, and that $\mu_q$ is as effective as the Lyapunov exponent \cite{ch1}. 
Multiparticle production in $\pi p$ and $Kp$ collisions in fixed target experiments at 250 GeV has been studied by use of erraticity analysis; it is found that at such low collision energy the erraticity behavior is dominated by purely statistical fluctuations \cite{ata}.
For light-ion collisions and  emulsion data at low energies  the erraticity index $\mu_q$ has  been determined with  results that cannot be used to shed any light on PT \cite{glg,dg}. More recently, it has been considered in model studies at LHC energies \cite{hy12,gup}. Now,  it is eminently suited for the real data at $\sqrt{s_{NN}}=2.76$ TeV to be analyzed.

In order to have a hadronization scheme that simulates critical transition, it is attempted in Ref.\ \cite{hy12} to build in two opposing subprocesses at each time step: one is contraction of dense regions that mimics confinement attraction, and the other is randomization by redistributing the $q$ and $\bar q$ in the dilute bins randomly throughout the \ep\ space. The former represents ordered motion and the latter disordered motion. Pionization occurs whenever a pair of $q$ and $\bar q$ are within an assigned small distance from each other. Applying erraticity analysis to the multiplicity fluctuations generated in that model results in the erraticity index for $q=4$ being $\mu_4\ ^< _\sim\ 2$ for the critical case, but twice larger for the no-contraction case \cite{hy12}.

\section{Transverse-momentum spectra at RHIC and LHC}

In the previous section we considered the multiplicity fluctuations in \ep\ space in small $\Delta p_T$ intervals at low \ppt\ in search for signatures of quark-hadron PT. Now, we consider the orthogonal problem of \ppt\ distributions over wide ranges of \ppt, averaged over $\eta$ and $\phi$. We summarize here only the approach based on the recombination model (RM), in which hadronization is treated by the recombination of thermal and shower partons \cite{hy03}. It is the effect of the increase of shower partons, as energy is increased from RHIC to LHC, that we shall discuss in the next section. 

Since our aim is primarily focused on the effects of minijets on the signature for PT, it is sufficient for us to restrict our discussion here to only the production of pions at midrapidity, i.e., $|\eta|<0.5$, with $\phi$  averaged over $2\pi$ for central collisions. Thus the \ppt\ distribution appears as a 1D problem with 
\bq
{dN_{\pi}\over \pt d\pt}={1\over p^0p_T} \int {dp_1\over p_1}{dp_2\over p_2} F(p_1,p_2) R_{\pi}(p_1,p_2,p_T)   \label{15}
\eq
where $p_1$ and $p_2$ are the transverse momenta of a quark and an antiquark that recombine to form a pion.  $R_{\pi}$ is the recombination function (RF), which includes a $\delta(p_1+p_2-p_T)$ function to conserve momentum. $F(p_1,p_2)$ is the invariant distribution of $q$ and $\bar q$ at late time when quark-hadron transition takes place. That transition is not referred to as PT because the RM is a microscopic description of hadron formation without collective dynamics, so it does not contain the macroscopic features discussed in the preceding section. As mentioned earlier, this is the orthogonal problem.

	The kernel of attention in the RM is the determination of $F(p_1,p_2)$. Without following through the evolution of the system as in hydrodynamics (which involves assumptions that are inconsistent with our premise that minijets can be important), $F(p_1,p_2)$ cannot be computed from the initial configuration at early time. The parton distributions due to hard and semihard jets can be calculated. The shower partons (S) are the fragmentation products of the hard and semiarid partons that emerge from the surface after their momenta are degraded by the medium they traverse. Without going into the details that can be found in Refs.\  \cite{hy03,hy04,hy10,hz1,zh}, let it simply be stated that the shower-parton distribution, denoted by ${\cal S}(p_i)$, can be calculated subject to a few adjustable parameters that are mainly associated with the momentum degradation process over a wide range of hard- and semihard-parton momenta, extending beyond the validity of pQCD on the low virtuality side. ${\cal S}(p_i)$ depends on collision energy, but for any given energy it can be determined by fitting the pion distribution at high \ppt\ using \ee(15) and the identification of $F(p_1,p_2)$ with SS recombination in a single jet. That is, if $\hat F_i(q)$ denotes the distribution of a hard or semihard parton of type $i$ with momentum $q$ upon emerging from the medium, and $S_i^j(p_1,q)$ is the unintegrated distribution of shower parton $j$ fragmented in vacuum from parton $i$, then the integrated distribution is
\bq
{\cal S}^j(p_1)=\int{dq\over q}\sum_i \hat F_i(q) S_i^j(p_1,q)   \label{16}
\eq
and the two-shower contribution to $F(p_1,p_2)$ in \ee(\ref{15}) is
\bq
F_{SS}^{j\bar j}(p_1,p_2)=\int{dq\over q}\sum_i\hat F_i(q)S_i^j(p_1,q)S_i^{\bar j}\left({p_2\over q-p_1},q\right) \ ,  \label{17}
\eq
which will be abbreviated by the symbolic notation $\cal SS$.  It is the degradation parameters in $\hat F_i(q) $ that are adjusted to fit the pion distribution at high \ppt\ where $\cal SS$ dominates. Once $\hat F_i(q)$ is determined phenomenologically this way, then ${\cal S}(p_1)$ is  known through \ee(16) even at low $p_1$ where thermal partons in the bulk medium cannot be neglected.

	The notion of thermal partons deserves extensive discussion, especially since thermal interaction in the context of PT discussed in the previous section has a slightly different meaning from what is to be considered here. Postponing that discussion until the next section, let us proceed with our brief summary of how the pion \ppt\ distribution is calculated in the RM. Minijets are clusters of particles that stand above the background, which consists of all the particles that the QGP transforms into. At the parton level just before hadronization the former are the shower partons, while the latter are called thermal partons, since they are the constituents of the equilibrated medium. With ${\cal T}(p_1)$ denoting the distribution of the thermal partons, the two-parton distribution $F(p_1,p_2)$ in \ee(15) is a sum of all possible pairings of $\cal T$ and $\cal S$, and can therefore be represented symbolically as 
\bq
F(p_1,p_2)={\cal TT+TS+SS}.  \label{18}
\eq
If minijets are rarely produced, then TT recombination is all that is needed to reproduce the data on pion spectrum.

	At RHIC the effects of ${\cal S}(p_2)$ at low $p_2$ are negligible even though it is dominant at high $p_2$, so TT recombination alone is sufficient to account for the low-\ppt\ behavior of $dN_\pi/\pt d\pt$, which is exponential. Thus, by adopting the form \cite{hy04} 
\bq
{\cal T}(p_1)=Cp_1e^{-p_1/T} \ ,   \label{19}
\eq
the pion distribution can be well fitted for $\pt<3$ GeV/c with
\bq
C=23.2\ {\rm GeV}^{-1} \ , \qquad\qquad	T=0.317\ {\rm GeV} \ .   \label{20}
\eq
These are phenomenological quantities determined from central Au-Au collisions at $\sqrt{s_{NN}}=200$ GeV. The ${\cal S}(p_1)$ distribution for the same collisions, when calculated according to \ee(16), is lower than ${\cal T}(p_1)$ for $p_1<2$ GeV/c, and higher above. Both ${\cal T}(p_1)$ and ${\cal S}(p_1)$ are shown in Fig.\ 19 of Ref.\ \cite{zh}.

	At LHC there are many more shower partons because of copious production of minijets at higher collision energy. The ratio ${\cal S}^{\rm LHC}(p_1)/{\cal S}^{\rm RHIC}(p_1)$ is shown in Fig.\ 6 in Ref.\ \cite{zh}, increasing from a value of $\sim$7 at $p_1\sim 0.5$ GeV/c to $\sim$30 at $p_1\sim 3$ GeV/c (and more at higher $p_1$) for Pb-Pb collisions at $\sqrt{s_{NN}}=2.76$ TeV. Such a large increase of ${\cal S}(p_1)$ cannot be accompanied by a similar increase of ${\cal T}(p_1)$, or even a milder increase, because the inclusive charged particle multiplicity per participant pair at midrapidity increases by only a factor of $\sim$2. In fact, the pion spectrum at LHC can be well reproduced over the whole range of $1<\pt<20$ GeV/c with ${\cal T}(p_1)$ remaining the same as at RHIC, specified by Eqs.\ (19) and (20) \cite{zh}. The ratio ${\cal S}(p_1)/{\cal T}(p_1)$ 
increases from $\sim$1 at $p_1\sim 0.5$ GeV/c to $\sim$$10^2$ at $p_1\sim 6$ GeV/c. With ${\cal S}(p_1)$ being approximately equal in magnitude to ${\cal T}(p_1)$ at low $p_1$, it becomes necessary to ask some questions on the effect of shower partons on observables related to quark-hadron PT.

\section{Thermal and Shower Partons}

In Sec.\ II we have discussed multiplicity fluctuations in the \ep\ space with the aim of finding possible signals of quark-hadron PT. In Sec.\ III we have reviewed the orthogonal problem of determining the \ppt\ distribution for the purpose of exposing the underlying quark distributions of the two types, thermal and shower. At LHC the two problems intersect at low \ppt\ in the region $0.2\ ^< _\sim\ p_T\ ^< _\sim\ 1$ GeV/c, since TS recombination dominates at $\pt>1$ GeV/c. At RHIC TS dominance does not occur until $\pt>3$ GeV/c, so there is a wider region below 3 GeV/c to examine the fluctuation patterns in \ep\ with minimal effects of minijets. However, small $\Delta p_T$ cuts still need to be made to minimize the overlap of spatial patterns, and at $\sqrt{s_{NN}}=200$ GeV there are not enough produced particles to populate the \ep\ space when $\Delta p_T$ is as narrow as 0.1 GeV/c. For that reason LHC offers for the first time realistic opportunity to study quark-hadron PT. Hereafter our attention will be focused on only the Pb-Pb collisions at LHC in the region $\pt<1$ GeV/c, although the ubiquitous shower partons are understood to arise from minijets that are due to semihard partons with momenta $q>3$ GeV/c. It should be recognized, however, that the shower parton distribution at low $p_1$ could not have been determined without studying the pion distribution at high \ppt, as discussed in Sec.\ III.

	The thermal partons are not derived from hydrodynamics, which would be inadequate in the presence of abundant shower partons if the parameters in it are determined by neglecting minijets. In the RM the parametrization of ${\cal T}(p_1)$ is adjusted to fit the RHIC data at low \ppt. It is the distribution of partons (quarks and antiquarks with gluons converted to $q\bar q$ pairs, ready for confinement transition) at the end of evolution of the medium. At midrapidity in central collisions for which there is no azimuthal anisotropy ${\cal T}(p_1)$ is the average transverse momentum distribution that contains no information about the system outside the narrow $\Delta\eta$ interval at $\eta=0$. Hence it is insensitive to the edges of the rapidity plateau in the fragmentation region of the leading particles. Indeed, it should be independent of the collision energy (provided that it is high enough to have a mid-$\eta$ region well separated from the edges) because the transition from deconfined quarks to confined quarks in hadrons occurs at such a late stage of the medium's life span that the local properties of the quarks retain no memory of the initial temperature or configuration \cite{zh}. That is analogous to water vapor condensing at $100^{\circ}$C independent of how hot it has previously been. Thus Eqs.\ (19) and (20) have been used for ${\cal T}(p_1)$ for both RHIC and LHC collisions, resulting in good agreement with the data on $\pi$ and $p$ spectra for all $\pt>1$ GeV/c; in fact, using higher values of $C$ and $T$ than those in \ee(20) for LHC is found to yield too many hadrons at all \ppt\ \cite{zh}.

	Despite phenomenological success, the universality of ${\cal T}(p_1)$ stimulates further inquiries on what constitutes the thermal partons. In the RM they are regarded as whatever partons there are in the neighbor of a shower parton S so that a recombination process can take place between S and any of its near neighbors. Nothing is asked about the origin of that neighbor T that is not another S. The shower parton S, whose distribution is ${\cal S}(p_1)$, is the in-vacuum fragmentation product of a hard or semihard parton that has emerged from the medium. While in the medium, the hard or semihard parton can undergo scattering, radiation and other energy-losing processes. The energy lost to the medium enhances the disordered motion of the soft partons in it. They are called the enhanced thermal partons in the neighborhood of the trajectory of the semihard parton in a series of studies of the ridge and azimuthal anisotropy as an alternative to the hydro explanation \cite{hz1,h5,chy,hz2}. Being phenomenologically parameterized, ${\cal T}(p_1)$ includes the enhanced thermal partons that are difficult to calculate. They evolve as more gluon radiation interacts with more medium partons until at late time the distinction between the original and in-medium shower partons becomes meaningless. They reach local equilibrium and are referred to as thermal partons in a generic sense. However, the choice of the word ``thermal" does not imply that they are precluded from collective interaction when the local temperature $T$ gets down to $T_c$.

	It is reasonable to ask why, if ${\cal S}(p_2)$ increases with collision energy, ${\cal T}(p_1)$ does not. In \ee(16) the part that depends on $\sqrt{s_{NN}}$ is the hard-parton distribution $f_i(k)$ at the point of creation before $k$ is degraded to $q$ and $f_i(k)$ evolves into $\hat F_i(q)$ \cite{hy10, zh2,zh}. Thus ${\cal S}(p_2)$ increases because more minijets are created at higher energies. Although more harder jets are also created at higher $q$, their rates of production are orders of magnitude lower compared to the minijets so their effects on ${\cal S}(p_2)$ at low $p_2$ are negligible. With the increase of minijets the enhanced thermal partons are also increased. However, the important point about ${\cal T}(p_1)$ is that if the medium is denser as a result of the additional in-medium shower partons, the quarks in it would remain deconfined until further expansion reduces the density enough to become ready for recombination. ${\cal T}(p_1)$ is always the distribution at the end of the expansion phase and has the same inverse slope $T$ from RHIC to LHC, no matter how many more partons are produced initially or during expansion. The formalism of RM is not precise at very low momenta, since pion spectrum at $\pt<1$ GeV/c can contain contributions from resonance decays that are not accounted for. Thus ${\cal T}(p_1)$ is not expected to be accurate at $p_1<0.5$ GeV/c, which is the region that may well be where more thermal partons are enhanced at higher energy due to the prolongation of momentum-degradation processes in the extended expansion phase. Above $\pt>1$ GeV/c the RM has succeeded in reproducing not only the pion distribution at LHC, as mentioned at the end of Sec.\ III, but also the spectra of proton (with the same $T$) and strange particles (at slightly higher $T_s$ for strange quarks) at all \ppt\ where particle species are identified \cite{zh}.

\section{Effects of Minijets on Multiplicity Fluctuations}

We have now arrived at the point where enough has been presented to allow a meaningful discussion about the interplay between fluctuation analysis and minijet effects. The region in momentum space that is relevant is $-1<\eta<+1,\ 0<\phi<2\pi$ and $0.2<\pt<1$ GeV/c. The \ppt\ region is lower than where the RM is formulated to treat reliably, but \ppt\ distribution is not of interest in the fluctuation analysis. What we know from preceding sections is that at $\pt>1$ GeV/c TS recombination is dominant at LHC and that the shower partons are developed outside the medium carrying little information about the collective behavior of the medium. At $\pt<1$ GeV/c the thermal partons contain the in-medium shower partons that are equilibrated with the soft partons in the expanding medium so that they are not distinguishable. Though they are named ÒthermalÓ, those partons are not just in random disordered motion, but can participate in ordered near-neighbor confinement interaction before hadronization.

	To find the signal for quark-hadron PT, the region $0.2<\pt<1$ GeV/c should be divided into narrow $\Delta p_T$ intervals, say 0.1 GeV/c wide, in order to minimize the overlap of spatial patterns in \ep\ generated in different time intervals on the assumption that there is a close correlation between \ppt\ and hadronization time $t$, i.e., larger \ppt\ at earlier $t$, smaller \ppt\ at later $t$. If that assumption turns out to be unrealistic, there is another method of analysis that will be discussed in the next section. An easy way to test that assumption is to check whether the signature, such as the scaling exponent $\nu$, is essentially the same in the $0.5<\pt<0.6$ GeV/c interval as in a larger region $0.4<\pt<0.8$ GeV/c without partition. If not, then the smaller interval is likely to reveal the finer structure at the cost of larger   errors.

	Since the method of analysis described in Sec.\ II is aimed at quantifying the nature of clusters of particles produced in heavy-ion collisions, the effects of minijets are of concern. High-\ppt\ jets are known to be characterized by clusters of particles that are identified in experimental searches as towers in lego plots. Those are jets with $\pt>100$ GeV/c.  They are so rarely produced at ${\cal O}(<10^{-10})$ that their effects on fluctuation analysis are totally negligible. Minijets with $E_J\sim 5$ GeV, say, are more copiously produced in heavy-ion collisions but are not extensively investigated by LHC experiments. Each such minijet fragments into a cluster of a small number of particles which, if concentrated within a small cone in \ep\ could contaminate the clustering effect due to critical behavior. However, a cluster of five or six of those particles have a spread of \ppt\ values, most of which are at low $p_T\ (\ ^< _\sim\ 1\ {\rm GeV/c})$ . Since the interval in which fluctuation analysis is done should be small, e.g., $\Delta\pt=0.1$ GeV/c, the average multiplicity per minijet in such small windows is much less than 1. The addition of one particle with minijet origin to the particles from the bulk medium does not contribute to cluster formation. Although an event may have numerous minijets, they are uncorrelated and are therefore randomly distributed throughout \ep. Thus, again, there is no reason to expect ordered clustering from all minijets within a small $\Delta\pt$ window. Besides, Poissonian fluctuation is filtered out by factorial moments.

	A concrete way to investigate the problem is to use an event generator that can reproduce the hadronic spectra and to examine the multiplicity fluctuations in narrow $\Delta\pt$ windows. That is exactly what Gupta and Sharma have done in \cite{gup} using the AMPT model \cite{bz}. That model has no collective dynamics to simulate critical behavior, but presumably has minijets. They have found negative intermittency that corresponds to the intermittency index $\varphi_q$ in \ee(3) being negative. Thus, AMPT does not generate any clusters that can be measured by factorial moments in small bins. A dedicated event generator designed to produce minijets would be good for trial here to elucidate the effect of minijets on multiplicity fluctuations.

\section{Signal for critical behavior by void analysis}

In the preceding section we have seen how the clustering of particles from minijets does not affect the fluctuation analysis by factorial moments, provided that small $\Delta\pt$ windows are used with the consequence that each cluster is broken up into several uncorrelated intervals. It is also hoped that making the small $\Delta\pt$ cuts prevents the overlap of spatial patterns of hadronic clusters created at different times. That hope is based on an approximate one-to-one correspondence between \ppt\ and emission time. However, such a simple relationship may not be realistic. One can imagine that there is a Gaussian distribution in \ppt\ at any given emission time and that the mean $\bar p_T$ may decrease with increasing time but the width can be broader than the $\Delta\pt$ windows. Then in each \ppt\ interval the spatial patterns can receive contributions from several neighboring emission times. In such a scenario it is no longer persuasive to argue that the pattern of hadrons in \ep\ in a $\Delta\pt$ window of each event in a heavy-ion collision corresponds to a 2D configuration simulated in the Ising model. When there is a quark-hadron PT, it is not clear whether $F_q$ calculated according to \ee(2) using real data can be interpreted by the same calculation in the Ising model, as done in \cite{cgh}.

	A way to circumvent the complication discussed above is to eliminate the need to count the number $n$ of hadrons in a bin, which has to be $n\ge q$ in order to contribute to $F_q$. If for a fixed $\Delta\pt$ the number $M$ of bins is large enough such that on average the total number of particles detected in that \ppt\ interval is about, say, half of $M$, then a random distribution of those particles in the \ep\ space would generate a random array of occupied and empty bins, with more empty than occupied ones because more than one particle can be in a common bin. If collective interaction is at play as in a PT, then the clustering effect would result in more empty bins. Moreover, a region of connected empty bins can vary in size, both within one event and from event to event. Let us call that region without hadrons a void, to be defined more precisely below. Focusing on the size variation of voids allows us to ignore how many particles are piled up in the non-empty bins. In that way the effects of minijets are minimized. Moreover, it is appealing to make use of the physical sense that if hadronization time is divided into many steps, then a void in one time step is more likely to be followed by some hadrons emitted in that region in the next step. That is because hadrons are not emitted uniformly from the cylindrical surface of the plasma, but in patches in a given time interval, and as the medium expands, the quarks below the surface in the void region  are subject to stronger confinement forces,  thus more likely to hadronize there in the next time interval. That kind of physics can be built into a model and be tested in the void analysis \cite{hzh1,hzh2}.

	On the Ising lattice there are spins pointing up or down at each site. Let hadron density in a cell be defined in proportion to the net spin of the cell (containing 4$\times$4 sites, say) if it is positive, but defined to be zero, if it is negative. A bin containing several cells is defined to be empty if its average density is below a threshold $\rho_0$; otherwise, occupied. A void is defined to be a region consisting of several empty bins that are connected by at least one common side between adjacent bins. It is the size of a void that is of interest, not the densities of the occupied bins. The reason is that at the critical point the void regions can have all sizes. The presence of a few particles here and there due to minijets makes a negligible effect on the void sizes, even if the threshold $\rho_0$ is low, let alone when $\rho_0$ is raised. The point is to suppress small fluctuations and concentrate on large patches of contiguous empty bins. An analogy is the study of the topography of a rough terrain by flooding it with water and measuring only the areas of the submerged regions at different water levels.

	Using $V_k$ to denote the total number of empty bins in the $k$th void, and defining $x_k=V_k/M$ to be the fraction of bins in the space occupied by $V_k$, where $M$ is the total number of bins, let us define
\bq
G_q={g_q\over g_1^q} \ ,  \qquad\qquad  g_q={1\over m} \sum_{k=1}^m x_k^q \ ,  \label{21}
\eq
where $m$ is the total number of voids in a configuration. We can consider two averages over all configurations (or events): one is just $\left< G_q\right>$, the other is 
\bq
S_q=\left< G_q \ln G_q\right> .   \label{22}
\eq
They have been simulated in the Ising model for various thresholds $\rho_0$ in Ref.\ \cite{hzh1} and found to have scaling behaviors
\bq
\left< G_q\right> \propto M^{\gamma_q}  \qquad\qquad  {\rm and}  \qquad\qquad S_q \propto M^{\sigma_q} \ .  \label{23}
\eq
	Moreover, $\gamma_q$ and $\sigma_q$ exhibit linear dependences on $q$
\bq
\gamma_q=c_0+c\ q \ , \qquad\qquad \sigma_q=s_0+s\ q .  \label{24}
\eq
	The values of $c$  and $s$ vary about $\pm10\%$ as functions of the threshold $\rho_0$, depending on $T\ ^< _\sim\ T_c$. They converge to approximately constant values independent of $T$ when $\rho_0$ is on the low end at about 8\% of the maximum density. Their values are
\bq
	c=0.8   \quad\qquad {\rm and}  \qquad\qquad  s=0.76 \ ,    \label{25}
\eq
which are numerical characterization of critical behavior by void analysis in the Ising model \cite{hzh1}.

	It is possible to go closer to quark-hadron PT in heavy-ion collisions by generating temporally integrated configurations, for which a range of time steps in the Ising simulation are allowed to populate a $\Delta\pt$ window according to a Gaussian distribution in \ppt\ \cite{hzh2}. Preference for void regions to be occupied in successive steps can also be introduced. The result again exhibits the scaling behavior of \ee(23) for a wide range of $\rho_0$. The slopes $c$ and $s$ in (24) turn out also to be very similar to those given above, i.e.,
\bq
	c=0.8\pm 0.05 \quad\qquad {\rm and}  \qquad\qquad s=0.8\pm 0.1 \ .   \label{26}
\eq
	At a critical point there is so much fluctuation that many bins are empty. The scaling behavior found about the voids is evidently a property of the second-order PT that is independent of the details of time evolution.
	
	The void analysis can readily be applied to the real data at LHC. Dividing the \ep\ space into a square lattice of M bins and counting only the {\it connected} empty bins whose hadron densities are lower than a threshold \rh\ (so as to calculate the void moments $G_q$) are easier to perform than calculating factorial moments $F_q$, since there are no horizontal average to determine first. It is important, however, to be sure that the empty bins are connected to form a void. Sharing a corner only is not a connection. The scaling behaviors of $\left<G_q\right>$ and $S_q$ have been found in the analyses of emulsion data at low energy \cite{mm} where the values of $c$ and $s$ are lower than those given in \ee(\ref{26}), which is not a surprise since PT is not expected. We expect the void analysis to be an effective tool to study the critical behavior at LHC, if it exists.
	
\section{Conclusion}
	
We have reviewed two approaches in the diagnostics of critical behavior in heavy-ion collisions and considered the effects of minijets. The first approach is to analyze multiplicity fluctuations, while the second is to do the complementary analysis on voids where no hadrons are formed. Scaling behaviors are found in both approaches, based on theoretical calculations in accordance with the Ginzburg-Landau theory of phase transition and Ising model simulations. The lack of models that specifically address the phase transition problem in heavy-ion collisions is an indication of the difficulty in incorporating collective dynamics in a quark-gluon system that is dilute and ready for confinement. Calling the process freeze-out avoids the confrontation with the complexity of the problem, which may be necessary to get quickly to the hadronic distributions that are observed, but not to gain insight into a very different realm of physics that has not been explored experimentally. There is, therefore, an urgent need for the existing data from LHC to be analyzed in the manner suggested here in the hope of finding some preliminary signs of quark-hadron PT that can stimulate a movement toward developing in-depth theories focused on criticality in nuclear collisions at high energy. Such theories may suggest new observables that can invigorate dedicated experiments to reveal new physics.
	
	Minijets and maxijets may initially seem like detractors in the search for genuine signature of clusters due to critical behavior. However, as we have seen, so long as the \ppt\ variable is divided into small $\Delta\pt$ intervals, the fragments of jets will not affect either the multiplicity or void analysis in those independent intervals. That can be tested just by applying the analyses to the data generated by any of the existing codes on heavy-ion collisions, or even $pA$ and $pp$ collisions.
	
The conventional wisdom in the nuclear community is that the quark-gluon system created in nuclear collisions at high energies from RHIC to LHC behaves like a fluid that flows according to hydrodynamics. The hydro treatment ignores the effects of minijets that are abundantly produced at LHC. The hard and semihard partons that are responsible for those jets lose energy while traversing the medium that takes some time, and the lost energy takes more time to thermalize the bulk partons. Thus, the assumption of rapid equilibration in the hydro formalism is not realistic, let alone the neglect of shower partons that dominate over the thermal partons. Does that mean that there is no thermal system to which one can apply the usual description of critical phenomenon, such as that of Ginzburg-Landau? This is where the results from analyses of the present LHC data can provide some answers. If there are signs of scaling behavior described here, they are evidences in support of quark-hadron PT, which in turn suggests that by the time of hadronization the quark-gluon system has reached local thermal equilibrium with temperature being a good characterization, even though not directly measurable. The stage would then be set for further investigation on how the system with turbulent beginning relaxes to a thermal system at the end that includes the shower partons as the relics of hard processes at early time. More pertinent to the problems discussed here is not so much how the system gets there, but what the detailed nature of the PT is, when quarks become bound to form hadrons in the semi-global perspective of a collective phenomenon.

 \section*{Acknowledgment}
	
This work relies heavily on previous collaborations with Z.\ Cao, C.\ B.\ Yang, Q.-H.\ Zhang, and L.\ Zhu, whose participation in this long journey is gratefully acknowledged.


 \end{document}